\documentclass[prl,twocolumn,showpacs,preprintnumbers,amsmath,amssymb,superscriptaddress]{revtex4-1}

\usepackage{graphicx}
\usepackage{dcolumn}
\usepackage{bm}
\usepackage{epsfig}
\usepackage{amsmath}
\usepackage{amssymb}
\usepackage{color}
\usepackage{bbm}
\usepackage[caption=false]{subfig}
\usepackage{placeins}
\usepackage{soul}

\usepackage[colorlinks=true,citecolor=blue,linkcolor=blue,urlcolor=blue]{hyperref}
\usepackage{cancel}

\def\equationautorefname~#1\null{%
	Eq.~#1\null
}
\def\figureautorefname~#1\null{%
	Fig.~#1\null
}

\begin{document}

\title{Tripartite Interactions Induced Strongly Correlated Quantum Emissions}
\author{Qian Bin}
\email{qianbin@scu.edu.cn}
\affiliation{College of Physics, Sichuan University, Chengdu 610065, China}

\author{Ying Wu}
\affiliation{School of Physics, Huazhong University of Science and Technology, Wuhan, 430074, China}

\author{Franco Nori}
\affiliation{Theoretical Quantum Physics Laboratory, Cluster for Pioneering Research, RIKEN, Wako-shi, Saitama 351-0198, Japan}
\affiliation{Quantum Information Physics Theory Research Team, Quantum Computing Center, RIKEN, Wakoshi, Saitama, 351-0198, Japan}
\affiliation{Physics Department, The University of Michigan, Ann Arbor, Michigan 48109-1040, USA}

\author{Xin-You L\"{u}}
\email{xinyoulu@hust.edu.cn}
\affiliation{School of Physics, Huazhong University of Science and Technology, Wuhan, 430074, China}

\date{\today}

\begin{abstract}

Efficient generation of multiquanta emission is crucial for quantum information processing but remains challenging due to its typical reliance on higher-order quantum processes. Here, we theoretically demonstrate strongly correlated photon–phonon emission enabled by direct tripartite interaction. This interaction facilitates the formation of high-order multiquanta states without more intermediate state transitions, thereby avoiding the suppressed transition rates associated with multiple sequential processes and substantially improving resonant transitions. As a result, high-efficiency strongly correlated  even-quanta emission  (e.g., two photons and two phonons) can be achieved in the presences of dissipation. Beyond that, we show that introducing two-photon dissipation enables strongly correlated odd-quanta emission (e.g., two photons and one phonon)  in the tripartite interaction system by  parity-protected suppression of single-photon loss  and reconstruction of higher-order multiquanta processes.  Our work extends multiquanta emission into the tripartite coupling regime and holds promising potential for applications in hybrid quantum networks.
\end{abstract}

\maketitle

Multiquanta physics, which investigates processes involving the simultaneous participation of multiple energy quanta,  is emerging as an important frontier in quantum science and technology, with broad applications in quantum communication\,\cite{Pan2012CL, Verma2019RC, Llewellyn2022DF},  quantum biodetection\,\cite{Chu2011HZ, Horton2013WK, Li2023CZ}, and quantum metrology\,\cite{Joo2011MS, Qin2023DZ,Deng2024LC} and lithography\,\cite{DAngelo2001CS, Gonzalez-Hernandez2023VB, Zyla2024Farsari}.  One of its important branches is multiquanta emission, in which multiple photons, phonons, or other excitations are emitted almost simultaneously\,\cite{Hayat2008GO, Munoz2014dVT,  Strekalov2014}. Such processes enable the generation of highly entangled states\,\cite{Munoz2015LT, Liu2025WS} and allow the creation of exotic quantum states inaccessible via single-quantum transitions\,\cite{Chu2018KY}. They not only reveal rich quantum statistical properties\,\cite{Munoz2014dVT, Casalengua2020CL} but also support advance applications in quantum precision measurement\,\cite{Deng2024LC} and the development of nonclassical light sources\,\cite{Hayat2007GO, Chang2016GM}. In recent years,  strongly correlated multiquanta emission has been extensively explored in diverse platforms, including semiconductor\,\cite{ Ota2011IK, Heinze2017ZS, Lu2019SL}, cavity\,\cite{Munoz2018LdV, Bin2020LL, Ren2021DX, Deng2021SY, Gou2022HW, Bin2024QW} and circuit\,\cite{Bin2021BW, Ma2021LL, Liu2023HT,Xiong2025BC} quantum electrodynamics (QED) systems, and mechanical resonators\,\cite{Dong 2019Li, Zou2022LL, Bin2024JW}.  However, these processes typically rely on engineered higher-order nonlinearities or multi-excitation resonances, which often suffer from weak transition rates and low emission rates, resulting in low efficiency. Achieving high-efficiency multiquanta emission therefore remains a significant challenge. 

Recently, direct tripartite interactions—in which three distinct degrees of freedom interact simultaneously without decomposing into pairwise processes—have been proposed and are attracting increasing attention\,\cite{Cotrufo2017Fiore, Kani2022ST, Wang2022YS, Zhou2022HL, Hei2023LP, Tang2024Deng}. 
In contrast to conventional pairwise couplings, which mediate interactions through sequential excitation pathways, direct tripartite interactions intrinsically link multiple modes at the single-quantum level, enabling the simultaneous creation or annihilation of excitations in different subsystems. As a result, they give rise to fundamentally new transition channels that are inaccessible in systems governed solely by pairwise interactions. Such tripartite couplings have been shown to enable a range of novel physical effects, including unconventional nonlinear dynamics\,\cite{Cotrufo2017Fiore, Kani2022ST, Wang2022YS}, enhanced control of multiparticle correlations\,\cite{Hei2023LP}, and the generation of complex nonclassical states involving multiple degrees of freedom\,\cite{Zhou2022HL,  Tang2024Deng}. Given their ability to fundamentally reshape the excitation transitions pathways, this approaches naturally raise the question is whether the direct tripartite interactions could be harnessed as powerful tools for achieving high-efficiency  multiquanta emission. Moreover, the crossover of the multiquanta physics and  tripartite interaction theory  remains largely unexplored, and may substantially advance the field of multiquantua physics.  

Here, we propose strongly correlated photon–phonon emission enabled by a tripartite interaction in a cavity–atom–mechanical system, where the motion of an atom trapped in a harmonic potential modulates the atom–cavity coupling, thus inducing a direct interaction among the atom, photons, and phonons at the single-quantum level\,\cite{ Kani2022ST}. This tripartite interaction enables the simultaneous excitation of photon and phonon modes and allows high-order multiquanta resonant excitations to occur without relying on cascaded intermediate transitions. By bypassing multiple sequential processes inherent to conventional pairwise coupling schemes, the proposed mechanism avoids the associated suppression of transition rates and thus substantially enhances multiquanta transition efficiencies. When combined with system dissipation, this leads to high-efficiency strongly correlated multiquanta emission, overcoming the limitations of conventional schemes based solely on pairwise interactions.  The resulting emission corresponds to even-quanta emission, such as one photon and one phonon or two photons and two phonons. Given that the parity of quantum states serves as a key resource for quantum error correction\,\cite{Terhal2015, Livingston2022BF} and entanglement purification\,\cite{Pan2001SB}, odd-quanta emission emerges as an equally important counterpart. Furthermore, we demonstrate that introducing two-photon dissipation---which permits only the emission of photon pairs---enables strongly correlated odd-quanta emission (e.g., two photons and one phonon). By suppressing single-photon decay channels due to parity-protected photon dissipation, this dissipation mechanism effectively reconstructs higher-order multiquanta processes and enables flexible control of multiquanta dynamics via tailored system–environment interactions\,\cite{Leghtas2015TP, Touzard2018GL, Lescanne2020VP, MBerdou2023MR}. Together, these results establish direct tripartite interactions as a powerful tool for realizing efficient multiquanta emission, opening new avenues for generating nonclassical states and exploring multiquanta physics in hybrid quantum systems.

\begin{figure}
  \centering
  \includegraphics[width=8.5cm]{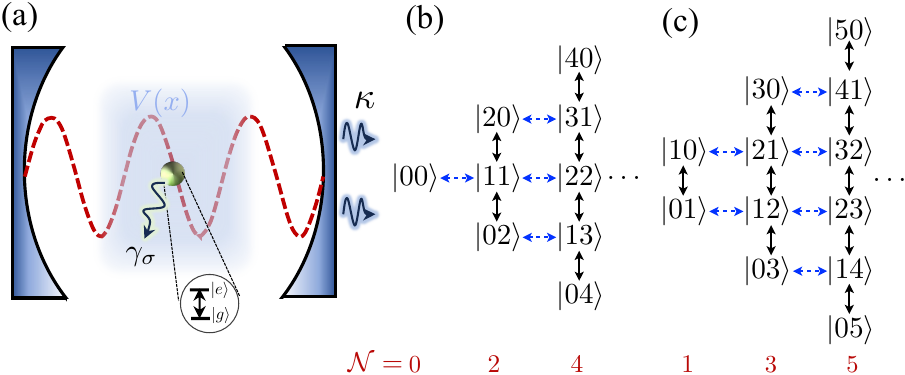}
  \caption{(a) Schematic of the system: a coherently driven two-level atom is trapped in a harmonic potential $V(x)$ at a node of the cavity field. The cavity photon mode ($a$), the atom ($\sigma$), and the mechanical center-of-mass phonon mode ($b$) are simultaneously coupled.  (b,c) State transitions within the (c) even-parity and (d) odd-parity subspaces of the two-boson-mode subsystem, where $\mathcal{N}=0,1,2\dots$ denotes the bosonic exaction number. 
\label{fig1}}
\end{figure}

\emph{Model and multiquanta resonances}.---As shown in Fig.\,\ref{fig1}(a), we consider a hybrid quantum system consisting of a two-level atom coupled to a cavity field, where the motion of the atom trapped in a harmonic potential modulates the atom–cavity coupling. The atom is coherently driven by a laser with amplitude $\Omega$ and frequency $\omega_L$.  Within the rotating-wave approximation, the system Hamiltonian in the basis $\{|e\rangle, |g\rangle\}$ reads  ($\hbar=1$)\,\cite{ Kani2022ST}
\begin{align}\label{eq01}
H=&\omega_{a} a^\dag a + \omega_{\sigma} \sigma_z/2 + \lambda_{a\sigma}\sin(kx)(a^\dag \sigma + a\sigma^\dag)\,\nonumber\\
& + \Omega (\sigma e^{i\omega_L t} +\sigma^\dag e^{-i\omega_L t} ),
\end{align}
where $a$ ($a^\dag$) is the annihilation (creation) operator of the cavity mode with resonance frequency $\omega_a$, $\sigma=|g\rangle \langle e|$ and $\sigma_z$ are the Pauli operators of the atom with transition frequency $\omega_{\sigma}$,  $k$ is the cavity field wave number, $x$ is the position of the atom in the cavity, and $\lambda_{a\sigma}$ is the coupling strength between the cavity mode and the atom. Since the atom is trapped in a harmonic potential its motion introduces motional degrees of freedom that must be taken into account. Expanding $\sin(kx)$ in Eq.\,(\ref{eq01}) to first order in the displacement around $x_0$, and choosing $x_0$ to coincide with a node of the cavity field $k x_0=n\pi$,  we obtain an interaction Hamiltonian that includes a direct tripartite coupling among the atom, cavity photons, and phonons at the single-quantum level $H_{\rm tri}= \lambda_{a b\sigma}(a^\dag \sigma + a\sigma^\dag) (b^\dag +b)$. The $b$ is the annihilation operator of mechanical mode with frequency $\omega_b$, and the tripartite interaction strength $\lambda_{ab \sigma} = \lambda_{a \sigma} k x_{\rm ZPM}$ is set by the zero-point fluctuation $x_{\rm ZPM}=\sqrt{\hbar/(2M\omega_b)}$, with $M$ is the effective mass of the atom.  This bare coupling is typically too weak to induce strong nonlinear effects, a parametric drive can be applied to the cavity mode to enhance the tripartite interaction strength\,\cite{Hei2023LP, Lu2015WJ, Qin2018ML, Qin204KM, supp}.  In the frame rotating at $\omega_L$, the effective system Hamiltonian can  be given by
\begin{align}\label{eq02}
H_{\rm eff}=&\Delta_{a} a^\dag a + \Delta_{\sigma} \sigma_z/2 + \omega_b b^\dag b + \lambda(a^\dag \sigma + a\sigma^\dag) (b^\dag +b)\,\nonumber\\
&+ \Omega (\sigma +\sigma^\dag ),
\end{align}
where $\Delta_{a, \sigma}$ are the frequency detunings of the laser from the cavity mode and atom, respectively, and $\lambda$ is the enhanced effective tripartite coupling strength.  

Considering that the laser resonantly drive the atom with $\Delta_{\sigma}=0$ and  in the strong driving limit $\Omega\sim\omega_b$, the laser field dresses the two-level atom and generates two new eigenstates of the driven-atom system, $|\pm\rangle =(|g\rangle\pm|e\rangle)/\sqrt{2}$, with corresponding eigenenergies $E_{|\pm\rangle}=\omega_{\sigma}\pm\Omega$\,\cite{Casalengua2020CL,Munoz2018LdV,Mollow1969,Cohen-Tannoudji1977Reynaud,Ulhaq2012WU}. Transforming the Hamiltonian Eq.\,(\ref{eq02}) into the dressed basis, where $\sigma_x\to\tilde{\sigma}_z$,  $\sigma_z\to\tilde{\sigma}_x$, and $\sigma_y\to-\tilde{\sigma}_y$,  we obtain
\begin{align}\label{eq03}
H_{\rm eff}'=&\Delta_{a} a^\dag a + \omega_b b^\dag b +\Omega\tilde{\sigma}_z + \frac{1}{2}\lambda[ (a^\dag b^\dag + a^\dag b + a b^\dag + ab)\tilde{\sigma}_z \,\nonumber\\
&+ (a^\dag b^\dag+a^\dag b-a b^\dag-ab)(\tilde{\sigma}^\dag-\tilde{\sigma})],
\end{align}
with $\tilde{\sigma}_z=|+\rangle\langle +| - |-\rangle \langle -|$ and $\tilde{\sigma}=|-\rangle\langle +|$. The effective Hamiltonian describes that the two-boson-mode subsystem preserves the parity symmetry\,\cite{Braak2011,Casanova2010RL}. The eigenstates of this two-mode subsystem can therefore be classified as even- and odd-parity states, corresponding to even and odd total excitation numbers, and form two independent, parity-conserving subspaces of the subsystem dynamics—namely, the even- and odd-parity subspaces, as shown in Figs.\,\ref{fig1}(b,c).  States prepared with a given parity evolve exclusively within the corresponding subspace, and transitions between the even- and odd-parity subspaces are strictly forbidden.

\begin{figure}
  \centering
  \includegraphics[width=8.7cm]{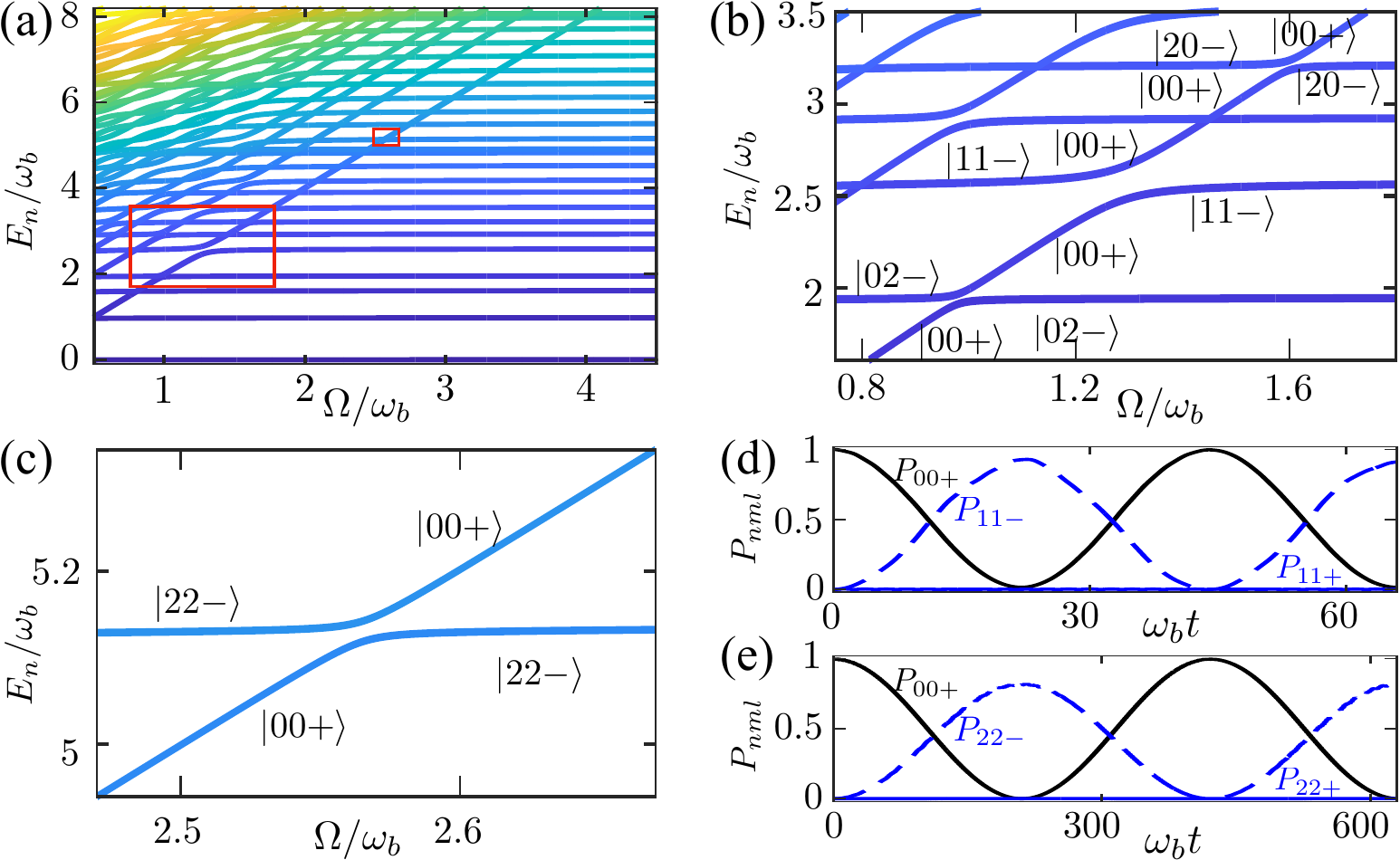}
  \caption{Energy spectrum of the Hamiltonian $H$ as a function of $\Omega/\omega_b$. (b,c) Enlarged views of the red-boxed regions in (a). (d,e) Time evolution of the state populations $P_{nml}=|\langle nml|\psi(t)\rangle|^2$ ($l=\pm$ and $n,m=0,1,2$) at multiquanta resonances in the absence of dissipation.  System parameters: $\Delta_{a}/\omega_b = 1.6$, $\lambda/\omega_b=0.15$, (d) $\Omega/\omega_b=1.3$, and (e) $\Omega/\omega_b=2.6$.
   \label{fig2}}
\end{figure}

By numerically diagonalizing the effective Hamiltonian in Eq.\,(\ref{eq03}), we obtain the eigenenergies $E$ and corresponding eigenstates. Figure\,\ref{fig2}(a) shows the engenergy differences  $E_n=E-E_0$ between the eigenenergies and the lowest energy state as a function of the laser driving amplitude $\Omega$.  Multiple regions of avoided level crossings can be observed, each indicating a resonant coupling between specific eigenstates. Importantly, these resonant couplings occur only between states within the same parity subspace of the two-boson-mode subsystem; avoided crossings do not appear between eigenstates of different parities. In particular, we focus on the resonant coupling between even-parity states. Figures\,\ref{fig2}(b) and \ref{fig2}(c) show enlarged views highlighting several avoided-crossings between the state $|00+\rangle$ and a family of even-excitation states $|n_a n_b -\rangle$, where $n_a, n_b=0,1,2\dots$ and $n_a+n_b$ is even. Each avoided crossing arises from the hybridization of the states $|00+\rangle$ and $|n_a n_b -\rangle$, with the resulting eigenstates well approximated by $(|00+\rangle\pm|n_a n_b -\rangle)/\sqrt{2}$ at the corresponding avoided-crossing points. As the driving amplitude $\Omega$ is tuned toward a given avoided-crossing points $\Omega\approx(n_a \Delta_{a} + n_b \omega_b)/2 $\,\cite{supp}, i.e., a multiquanta resonant position,  the state $|00+\rangle$ can resonantly couple to the even-exaction  states $|n_a n_b -\rangle$. In this way,  the system can be coherently driven from $|00+\rangle$ into higher even-exaction states within the same parity subspace of the two-boson-mode subsystem. These resonant transitions can also be verified by the multiquanta super-Rabi oscillations  $|00+\rangle \leftrightarrow |11-\rangle$ and $|00+\rangle \leftrightarrow |22-\rangle$, shown in Figs.\,\ref{fig2}(d) and \ref{fig2}(e), respectively. 

\begin{figure}
  \centering
  \includegraphics[width=8.7cm]{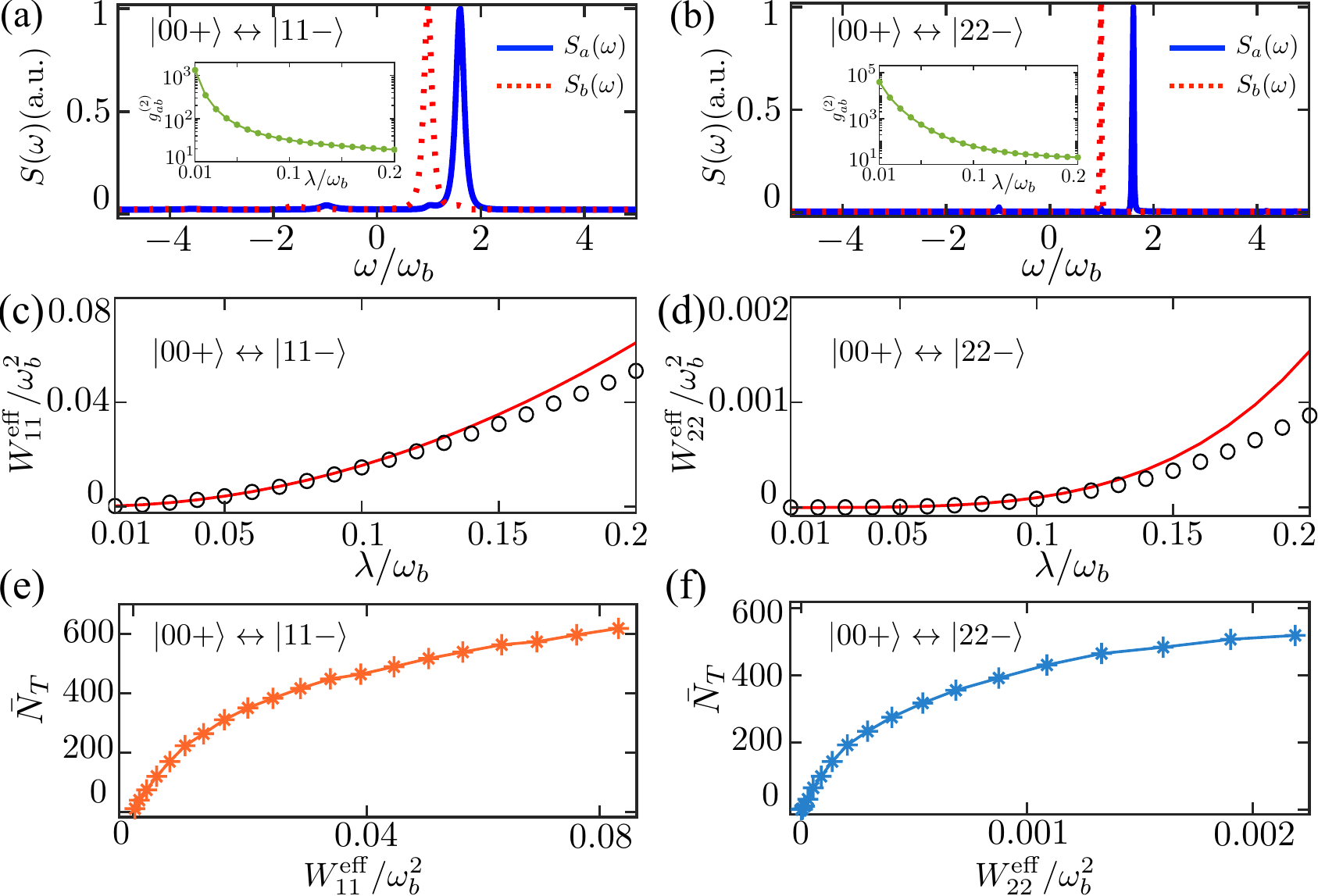}
  \caption{ (a,b) Photon and phonon emission spectra  $S_{a,b}$ when $\lambda/\omega_b=0.15$. Insets: equal-time second-order cross correlations $g_{ab}^{(2)}$ versus $\lambda/\omega_b$. (c,d) Effective multiquanta transition rates $W_{\rm 11,22}^{\rm eff}/\omega_b^2$ versus $\lambda/\omega_b$, obtained from full-numerical simulations (black circles) and analytical solutions  (red curves).  (e,f) Average number  of multiquanta emission events $\bar{N}_T$ versus $W_{\rm 11,22}^{\rm eff}/\omega_b^2$ (from full-numerical simulations) over a time window $T$. The left and right columns corresponds to  multiquanta resonances at $\Omega/\omega_b\approx1.3$ and $\Omega/\omega_b\approx2.6$, respectively. Other system parameters: $\Delta_{a}/\omega_b = 1.6$, (a,e)$\kappa_a=\kappa_b=10\gamma=0.25\omega_b$,  (b,f)$\kappa_a=\kappa_b=10\gamma=0.025\omega_b$, (e) $\omega_bT=10^5$, and (f) $\omega_bT=10^6$.
\label{fig3}}
\end{figure}

\emph{Tripartite interaction induces strongly correlated even-quanta emissions}.---The resonant excitation of multiquanta states is a crucial step in realizing multiquanta emission. In quantum emission processes, system dissipation serves as the trigger for photon and  phonon releases. The occurrence of multiquanta emission can be revealed in the emission spectra by solving the master equation 
\begin{align}\label{eq04}
\frac{d}{dt}\rho = i[\rho, H_{\rm eff}]+\kappa_{a}\mathcal{L}[a]+\kappa_b\mathcal{L}[b]+\gamma\mathcal{L}[\sigma],
\end{align}
where $\mathcal{L}[\mathcal{O}]=(2O\rho o^\dag - \rho O^\dag O -O^\dag O \rho)/2$ is the Lindblad superoperator, $\gamma$ is the atomic spontaneous emission rate, and $\kappa_a$ and $\kappa_b$ denote the photon and phonon decay rates, respectively. In Figs.\,\ref{fig3}(a,b), we show the photon emission spectra $S_{a}= \lim_{t\to\infty}\int d \tau \langle a^\dag (t) a(t+\tau)\rangle e^{i\omega t}$ and the phonon emission spectra $S_{b}= \lim_{t\to\infty}\int d \tau \langle b^\dag (t) b(t+\tau)\rangle e^{i\omega t}$ at the two multiquanta resonances, $|00+\rangle \leftrightarrow |11-\rangle$ and $|00+\rangle \leftrightarrow |22-\rangle$. The emitted photons correspond to squeezed photons with  frequency $\omega_{e,a}=\omega+\omega_{L}$, arising from the squeezing transformation and the rotating frame defined by the driving frequency in the effective Hamiltonian, while the emitted phonon have frequency $\omega_{e,b}=\omega$. Hybrid multiquanta emission is clearly observed in the spectra, whit sharp peaks at $\omega=\Delta_{a}$ in the photon spectra and at $\omega=\omega_b$ in the phonon spectra, corresponding to the cavity-driving detuning and the photon mode frequency, respectively. Notably, these emissions are exclusively even-quanta because only even-excitation states are resonantly excited. They exhibit strong photon-phonon correlations, with the equal-time second-order cross correlations $g_{ab}^{(2)}\gg 1$ revealing pronounced bunching between emitted photons and phonons at both two multiquanta resonances, for a wide range of $\lambda/\omega_b$, as shown in the insets of Figs.\,\ref{fig3}(a,b).

Although dissipation enables the releases of strongly correlated photons and  phonons, excessively strong dissipation suppresses multiquanta emission: if the decay rate dominates over the multiquanta transition rate, the system cannot be fully excited to the target multiquanta state before relaxing to the vacuum. Enhancing the multiquanta resonant transition rate is therefore essential. In general, the effective  transition rate between an initial sate $|i\rangle$ and a finial state $|f\rangle$ can be written as $W^{\rm eff} =2\pi |V^{\rm eff}|^2 $, where $V^{\rm eff}$ denotes the effective coupling strength between $|i\rangle$ and $|f\rangle$ \,\cite{supp}. Typically, $V^{\rm eff}\propto\lambda^n\Omega^m$, where $\lambda$ and $\Omega$ are the interaction strength and driving amplitude of the system Hamiltonian, respectively,  and $(n+m)-1$ corresponds to the number of intermediate state involved in a given transition pathway\,\cite{Munoz2014dVT, Ma2015Law, Garziano2016MS}. Consequently, an increasing number of intermediate steps leads to a smaller effective coupling strength and thus a reduced transition rate. However, conventional multiquanta emission processes generally rely on higher-order transitions involving multiple intermediate states.

In our proposal, for the multiquanta resonance $|00+\rangle \leftrightarrow |11-\rangle$, the transition occurs directly via the tripartite interaction without intermediate states. This interaction enables the simultaneous creation of a photon and a phonon through $a^\dag b^\dag \tilde{\sigma}$, driving $|00+\rangle \rightarrow |11-\rangle$. This direct pathway substantially enhances the effective transition rate compared with conventional pairwise coupling schemes, where the excitation typically proceeds through at least one intermediate state. Similarly, for the higher-order multiquanta resonance $|00+\rangle \leftrightarrow |22-\rangle$, the transition requires only a single intermediate state in the tripartite interaction system, e.g., $|00+\rangle \rightarrow |11\pm\rangle \rightarrow |22-\rangle$ mediated by $a^\dag b^\dag \tilde{\sigma}^\dag(\tilde{\sigma} )$. This one-intermediate-state pathway still yields a significantly enhanced transition rate compared with traditional pairwise coupling schemes, which typically require at least three intermediate processes to reach $|22-\rangle$. To quantitatively verify this qualitative picture, we derive analytical expressions for the effective multiquanta  transition rates using perturbation theory. The rates for $|00+\rangle \rightarrow |11-\rangle$ and $|00+\rangle \rightarrow |22-\rangle$ are $W_{11}^{\rm eff} =2\pi |V_{11}^{\rm eff}|\approx\pi \lambda^2/2$ and $W_{22}^{\rm eff} =2\pi |V_{22}^{\rm eff}| \approx \pi\lambda^4[1/(2\Omega-\Delta_{ap}-\omega_b)+1/(\Delta_{ap}+\omega_b)]^2/2$, respectively\,\cite{supp}. Figures \ref{fig3}(c) and \ref{fig3}(d) show excellent agreement between the analytical results and full numerical simulations, confirming the validity of our approximation.   Furthermore, in Figs.\,\ref{fig3}(e,f) we show the average number of strongly correlated multiquanta emission events $\bar{N}_T$ within a time window $T$ versus the effective multiquanta transition rate $W_{\rm 11,22}^{\rm eff}/\omega_b^2$, obtained from full numerical simulations with $20$ random realizations. As the effective multiquanta transition rate increases, the number of correlated emission events within a fixed time window increases correspondingly. These results demonstrate that direct tripartite interaction enhances strongly correlated multiquanta emission by accelerating resonant transitions through the reduction of intermediate processes.

\begin{figure}
  \centering
  \includegraphics[width=8.7cm]{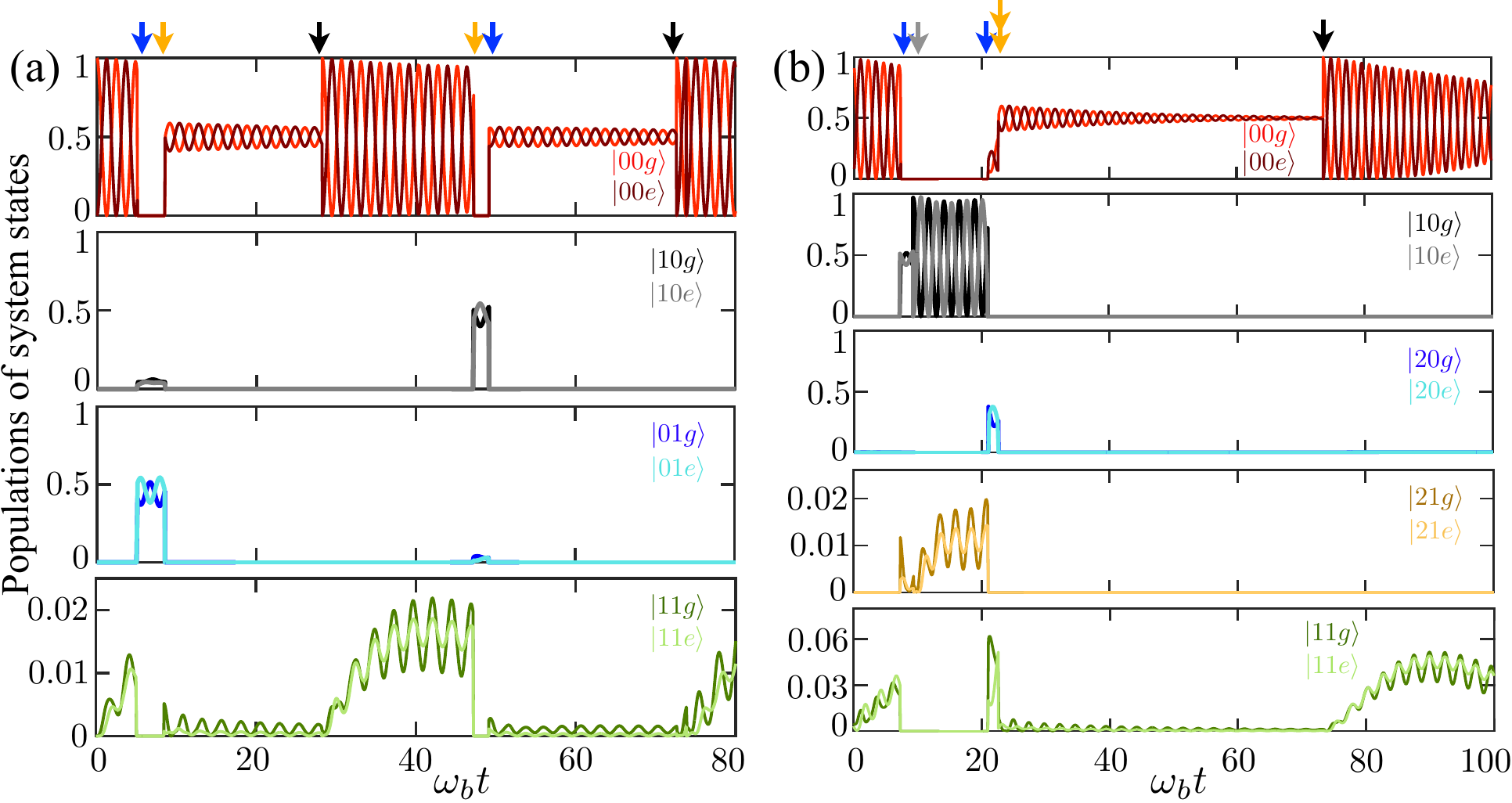}
  \caption{A tiny fraction of quantum trajectory in the multiquanta emission regimes of  single photons and single phonons  $|11-\rangle \to |00+\rangle $ represented by  population dynamics of different states for (a) single-photon dissipation case and (b) two-photon dissipation case. A (two) yellow (blue) arrow corresponding to a (two) single photon (phonon) emission, and the black arrow represents the atom decay.  System parameters: $\Delta_{a}/\omega_b = 1.6$, $\lambda/\omega_b=0.15$, $\kappa_a=\kappa_{a^2}=\kappa_b=10\gamma=0.25\omega_b$, and  $\Omega/\omega_b=1.3$.
  \label{fig4}}
\end{figure}

\emph{Strongly correlated odd-quanta emission enabled by two-photon dissipation}.---The parity of quantum states constitutes a fundamental degree of freedom and thus serves as a key resource for quantum error correction\,\cite{Terhal2015, Livingston2022BF} and entanglement purification\,\cite{Pan2001SB}. This makes strongly correlated odd-quanta emission, in addition to its even-quanta counterpart, equally significant.  Two-photon dissipation can induce such higher-order odd-quanta emission emission by modifying the lower-order emission process itself, e.g., for resonance $|00+\rangle \leftrightarrow|11-\rangle$.   Single-photon dissipation inherently triggers emission events one photon at a time, whereas two-photon dissipation only enables the simultaneous emission of photon pairs due to parity-protected photon dissipation. The dynamics of two-photon dissipation can be represented by replacing $\mathcal{L}[{a}]$ and $\kappa_{a}$ by $\mathcal{L}[{a^2}]$ and $\kappa_{a^2}$ in the master equation Eq.\,(\ref{eq04}), respectively, where $\kappa_{a^2}$ is the two-photon dissipation rate. Two-photon dissipation  has been experimentally realized through engineered environmental coupling in superconducting circuits\,\cite{Leghtas2015TP, Touzard2018GL, Lescanne2020VP, MBerdou2023MR}. By selectively suppressing single-photon loss channels, this process offers new possibilities for tailoring system dynamics\,\cite{Malekakhlagh2019Rodriguez, Roberts2021LC, Zhou2022ZL} and stabilizing nonclassical states\,\cite{Roberts2020Clerk, Munoz2021FS, Xu2023ZW}.

In Fig.\,\ref{fig4}, we present photon-phonon emission processes for the two photon-dissipation cases, obtained from quantum-trajectory simulations based on state populations.  Initially, the system mainly occupies a superposition state of the vacuum state $|00+\rangle$ and the photon-magnon state $|11-\rangle$. In single-photon dissipation case, a single photon and a single phonon are  emitted sequentially within a short time window, triggered by photon and phonon dissipation channels, respectively. The corresponding emission pathway can be described as $|11-\rangle \to |10-\rangle(|01-\rangle) \to |00-\rangle$.  After this strongly bunched photon-phonon emission and the subsequent atomic relaxation, the state $|11-\rangle$ is reconstructed by continuous driving, enabling repeated multiquanta emission cycles. As shown in Fig.\,\ref{fig4}(a), two photon-phonon emission events occur, and the order in which the photon and phonon are emitted is stochastic. In contrast, under two-photon dissipation,  single-photon emission cannot directly trigged. Instead,  after the emission of a single phonon and atomic relaxation, continuous driving re-excites the system to a higher-order multiquanta state $|21-\rangle$, which is inaccessible in the absence of dissipation due to parity symmetry constraints of the two-boson-mode subsystem. Once this state is populated, a second phonon and a photon pair are subsequently emitted through single-phonon and two-photon dissipation channels, respectively. The photon pair and the second phonon exhibit strongly bunching because they are emitted within a short time window, analogous to the single-photon dissipation case. In contrast, they show weak correlation with the first single phonon, since re-excited time required to reach the state $|21-\rangle$ exceeds the phonon lifetime. This overall emission process can be summarized as $|11-\rangle  \to |10-\rangle \to |10+\rangle \to |21-\rangle \to |20\rangle (|01-\rangle)\to |00-\rangle$,  corresponding to one phonon emission followed by strongly correlated  emission of a photon pair and an additional phonon,  as shown in Fig.\,\ref{fig4}(b).

\emph{Conclusion}.---We have shown that direct photon–phonon–atom tripartite interaction provides an efficient mechanism for generating  enhanced  multiquanta emission in a cavity–atom–mechanical system. By directly coupling photon and phonon excitations at the single-quantum level, the tripartite interaction enables resonant multiquanta transitions through reduced intermediate-state pathways, leading to enhanced transition efficiency and strongly correlated even-quanta emission. Furthermore, engineered two-photon dissipation reshapes the emission dynamics by suppressing single-photon loss channels and promoting higher-order correlated emission processes. This proposal provides a route for odd-quanta emission, tailoring emission properties in hybrid quantum platforms,  paving a promising way for developing controllable quantum sources for on-chip quantum communication and hybrid quantum network  applications\,\cite{Kimble2015, Dong2015WW, Riedinger2016HN, Wehner2018EH}.

\emph{Acknowledgments}.---This work is supported by the National Science Fund for Distinguished Young Scholars of China (Grant No.\,2425502), the National Natural Science Foundation of China (Grant No.\,12574397), the Fundamental Research Funds for the Central Universities (Grant No.\,2024BRA001), and the Sichuan Science and Technology Program (Grant No.\,2025ZNSFSC0057). F. N. is supported in part by Nippon Telegraph and Telephone Corporation (NTT) Research, the Japan Science and Technology Agency (JST) [via the Quantum Leap Flagship Program (Q-LEAP), and the Moonshot R$\&$D Grant No. JPMJMS2061], the Asian Office of Aerospace Research and Development (AOARD) (via Grant No. FA2386-20-1-4069), and the Office of Naval Research (ONR) (via Grant No. N62909-23-1-2074).

\onecolumngrid
\clearpage
\setcounter{equation}{0}
\setcounter{figure}{0}
\setcounter{table}{0}
\setcounter{page}{6}
\setcounter{section}{0}
\makeatletter
\renewcommand{\theequation}{S\arabic{equation}}
\renewcommand{\thefigure}{S\arabic{figure}}
\renewcommand{\bibnumfmt}[1]{[S#1]}
\renewcommand{\citenumfont}[1]{S#1}
\begin{center}
        \textbf{Supplemental Material for ``Tripartite Interactions Induced Strongly Correlated Quantum Emissions"}
\end{center}
\date{\today}
	
\title{Tripartite Interactions Induced Strongly Correlated Quantum Emissions}
\author{Qian Bin}
\email{qianbin@scu.edu.cn}
\affiliation{College of Physics, Sichuan University, Chengdu 610065, China}

\author{Ying Wu}
\affiliation{School of Physics, Huazhong University of Science and Technology, Wuhan, 430074, China}

\author{Franco Nori}
\affiliation{Theoretical Quantum Physics Laboratory, Cluster for Pioneering Research, RIKEN, Wako-shi, Saitama 351-0198, Japan}
\affiliation{Quantum Information Physics Theory Research Team, Quantum Computing Center, RIKEN, Wakoshi, Saitama, 351-0198, Japan}
\affiliation{Physics Department, The University of Michigan, Ann Arbor, Michigan 48109-1040, USA}

\author{Xin-You L\"{u}}
\email{xinyoulu@hust.edu.cn}
\affiliation{School of Physics, Huazhong University of Science and Technology, Wuhan, 430074, China}

\date{\today}
\maketitle
This supplemental material contains four parts. In Sec.\,I,  we provide a detailed derivation of the effective Hamiltonian used in the main text. In Sec.\,II, we provide a detailed derivation of the multiquanta resonance conditions in the Mollow regime. In Sec.\,III,  we derive the multiquanta effective transition rates via perturbation theory.

\section{I.\,Derivation of the effective Hamiltonian}\label{section1}
The Hamiltonian of the system, expressed in the basis $\{|e\rangle, |g\rangle\}$ ($\hbar=1$)
\begin{align}\label{eqI01}
H=&\omega_a a^\dag a + \frac{1}{2}\omega_{\sigma}\sigma_z + \lambda_{a\sigma}\sin(kx)(a^\dag \sigma + a \sigma^\dag) + \Omega(\sigma e^{i\omega_L t}+\sigma^\dag e^{-i\omega_L t}).
 \end{align}
Expanding the position-dependent coupling $\sin (kx)$ in Eq.\,(\ref{eqI01}) to first order in the displacement around $x_0$, with $k x_0=n\pi$, we obtain  
\begin{align}\label{eqI02}
H=&\omega_a a^\dag a + \frac{1}{2}\omega_{\sigma}\sigma_z + \omega_b b^\dag b + \lambda_{ab\sigma}(a^\dag \sigma + a \sigma^\dag) (b^\dag +b)+ \Omega(\sigma e^{i\omega_L t}+\sigma^\dag e^{-i\omega_L t}),
 \end{align}
where $b$ is the annihilation operator of the mechanical mode describing the atomic center-of-mass oscillation with trap frequency $\omega_b$. The tripartite interaction strength is given by $\lambda_{ab\sigma}=\lambda_{a\sigma} k x_{\rm ZPM}$, with $x_{\rm ZPM}=\sqrt{\hbar/(2M\omega_b)}$ denoting the zero-point fluctuation and $M$ the effective mass of the atom. Since this bare coupling is typically weak due to the small zero-point motion, we introduce a parametric drive on the cavity mode to enhance the tripartite interaction strength. The Hamiltonian then becomes\,\cite{Lu2015WJ, Qin2018ML, Qin204KM, Hei2023LP}
\begin{align}\label{eqI03}
H=&\omega_a a^\dag a + \frac{1}{2}\omega_{\sigma}\sigma_z + \omega_b b^\dag b + \lambda_{ab\sigma}(a^\dag \sigma + a \sigma^\dag) (b^\dag +b) + \Omega(\sigma e^{i\omega_L t}+\sigma^\dag e^{-i\omega_L t}) + \Omega_p (a^{\dag 2} e^{-i\omega_p t}+a^2 e^{i\omega_p t}),
 \end{align}
where $\omega_L=\omega_p/2$. In the rotating framework of driving frequency $\omega_L$,  Eq.\,(\ref{eqI03}) becomes
\begin{align}\label{eqI04}
H&=(\omega_a-{\omega_L}) a^\dag a +\frac{1}{2}(\omega_{\sigma}-{\omega_L})\sigma_z+\omega_b b^\dag b+\lambda_{ab\sigma} (a^\dag \sigma+a\sigma^\dag)(b^\dag+b)+ \Omega(\sigma +\sigma^\dag ) +\Omega_p(a^{\dag 2}+a^2)\nonumber\\
&=\Delta_{aL} a^\dag a +\frac{1}{2}\Delta_{\sigma L}\sigma_z+\omega_b b^\dag b+\lambda_{ab\sigma} (a^\dag \sigma+a\sigma^\dag)(b^\dag+b) + \Omega(\sigma +\sigma^\dag ) + \Omega_p(a^{\dag 2}+a^2),
\end{align}
where $\Delta_{aL}=\omega_a-{\omega_L}$ and $\Delta_{\sigma L}=\omega_{\sigma}-{\omega_L}$. We next apply a unitary squeezing transformation 
\begin{align}\label{eqI04-1}
U=\exp[r (a^{\dag 2}-a^2)],
\end{align}
with the squeezing parameter determined by 
\begin{align}\label{eqI04-2}
\tanh(4r)=2\Omega/\Delta_{aL}.
\end{align}
Neglecting constant energy shifts, the transformed Hamiltonian becomes
\begin{align}\label{eqI05}
H&=\frac{1}{2}\Delta_{\sigma L}\sigma_z+\omega_b b^\dag b-\frac{1}{2}\Delta_{aL} \sinh(4r_s)(a^{\dag 2}+a^2)+\frac{1}{2}\Delta_{aL} \cosh(4r_s)(2a^{\dag}a+1 )-\frac{1}{2}-\Omega\sinh(4r_s)(2a^{\dag}a+1)\nonumber\\
&~~~~-\lambda_{ab\sigma}\sinh(2r_s)(a^\dag \sigma^\dag+a\sigma)(b^\dag+b)+\lambda_{ab\sigma}\cosh(2r_s)(a^\dag \sigma+a\sigma^\dag)(b^\dag+b)+\Omega\cosh(4r_s)(a^{\dag 2}+a^2)\,\nonumber\\
&=\sqrt{\Delta_{aL}^2-4\Omega_p^2}a^{\dag}a+\frac{1}{2}\Delta_{\sigma L}\sigma_z+\omega_b b^\dag b-\lambda_{ab\sigma}\sinh(2r)(a^\dag \sigma^\dag+a\sigma)(b^\dag+b)\,\nonumber\\
&~~~~+\lambda_{ab\sigma}\cosh(2r)(a^\dag \sigma+a\sigma^\dag)(b^\dag+b) + \Omega(\sigma +\sigma^\dag )\,\nonumber\\
&=\Delta_{a}a^{\dag}a+\frac{1}{2}\Delta_{\sigma }\sigma_z+\omega_b b^\dag b+(\lambda'a^\dag \sigma^\dag+{\lambda'}^\ast a\sigma)(b^\dag+b)+\lambda(a^\dag \sigma+a\sigma^\dag)(b^\dag+b) + \Omega(\sigma +\sigma^\dag ),
\end{align}
where $\Delta_{a}=\sqrt{\Delta_{aL}^2-4\Omega_p^2}$, $\Delta_{\sigma}=\Delta_{\sigma L}$,  $\lambda=\lambda_{ab\sigma}\cosh(2r)$, and $\lambda'=-\lambda_{ab\sigma}\sinh(2r)$. The counter-rotating terms $a^\dag \sigma^\dag$ and $a\sigma$ describe  the non-conserving of excitation number and can be neglected under the condition $|\lambda'|/(\Delta_a+\Delta_{\sigma})\ll1$, corresponding to the rotating-wave approximation. The effective Hamiltonian is then given by
\begin{align}\label{eqI06}
H_{\rm eff}&=\Delta_{a}a^{\dag}a+\frac{1}{2}\Delta_{\sigma }\sigma_z+\omega_b b^\dag b+\lambda(a^\dag \sigma+a\sigma^\dag)(b^\dag+b) + \Omega(\sigma +\sigma^\dag ),
\end{align}
where $a$ denotes the annihilation operator of the squeezed cavity mode. In the strong-squeezing regime $2r\gg 1$, the effective atom-cavity coupling strength  can be significantly enhanced as 
\begin{align}\label{eqI07}
\lambda=\lambda_{ab\sigma}\cosh(2r)\approx\lambda_{ab\sigma}e^{2r}/2. 
\end{align}
\section{Derivation of multiquanta resonance conditions within the Mollow regime}
In the strong driving regime, the laser field dresses the two-level atom, and the driven-atom subsystem forms two new eigenstates that are quantum superposition of the bare states $\{|e\rangle, |g\rangle\}$. In the frame rotating at the driving frequency $\omega_L$, the  Hamiltonian of the strongly driven two-level atom  is given by
\begin{align}\label{eqII01}
&H_{\sigma}=\frac{1}{2}\Delta_{\sigma}\sigma_z + \Omega(\sigma+\sigma^\dag),
\end{align}
with the corresponding dressed eigenstates
\begin{align}\label{eqII02}
&|+\rangle=c_{+}|g\rangle + c_{-}|e\rangle,~~~~~~|-\rangle=c_{-}|g\rangle - c_{+}|e\rangle,
\end{align}
where the coefficients are
\begin{align}\label{eqII03}
&c_{\pm}=\sqrt{\frac{2\Omega^2}{(\Delta_{\sigma}^2+4\Omega^2) \pm \Delta_{\sigma}\sqrt{\Delta_{\sigma}^2+4\Omega^2}}},
\end{align}
satisfying  $|c_{+}|^2 + |c_{-}|^2=1$. The corresponding eigenenergies are
\begin{align}\label{eqII04}
&E_{|\pm\rangle}=\pm\frac{1}{2}\sqrt{\Delta_{\sigma}^2 + 4\Omega^2}.
\end{align}

For resonant driving, $\Delta_{\sigma}=\omega_{\sigma}-\omega_{L}=0$, the dressed eigenstates reduce to 
\begin{align}\label{eqII04-1}
|\pm\rangle=(|g\rangle \pm |e\rangle)/\sqrt{2},
\end{align}
with corresponding eigenenergies $E_{\pm}=\pm \Omega$. Transforming the effective Hamiltonian in Eq.(\ref{eqI06}) into the dressed-state basis $\{|+\rangle, |-\rangle\}$, 
\begin{align}\label{eqII04-2}
 \sigma_x\to\tilde{\sigma}_z, ~~\sigma_z\to\tilde{\sigma}_x,  ~~\sigma_y\to-\tilde{\sigma}_y. 
 \end{align}
 As a result,  Eq.(\ref{eqI06}) becomes
\begin{align}\label{eqII05}
H_{\rm eff}'&=\Delta_{a}a^{\dag}a+\omega_b b^\dag b+\Omega\tilde{\sigma}_z+\frac{1}{2}\lambda[(a^\dag b^\dag + a^\dag b + a b^\dag + ab)\tilde{\sigma}_z + (a^\dag b^\dag + a^\dag b - a b^\dag - ab)(\tilde{\sigma}^\dag - \tilde{\sigma})],
\end{align}
where  $\tilde{\sigma}_z=|+\rangle \langle + | - |-\rangle \langle - |$ and $\tilde{\sigma}=|-\rangle \langle + |$. Together with the photon and phonon modes, and neglecting the influence of the tripartite interaction on the energy structure, the eigenenergies of the system Hamiltonian can be given by
\begin{align}\label{eqII06}
&E_{|n_a n_b\pm\rangle}=n_a \Delta_a + n_b \omega_b \pm\Omega,
\end{align}
where $n_a,n_b=0,1,2,\dots$. To realize multiquanta resonant transitions, the total energy of $n_a$ photons and $n_b$ phonons matches the energy difference between the dressed states $|+\rangle$ and $|-\rangle$, i.e., $E_{|00+\rangle}=E_{|n_a n_b -\rangle}$, then 
\begin{align}\label{eqII07}
&n_a \Delta_a + n_b \omega_b - 2\Omega\approx0,
\end{align}
thus the driving amplitude $\Omega\approx(n_a \Delta_a + n_b \omega_b)/2$. Under this resonance condition, the initial state $|00+\rangle$ can be resonantly excited to the state $|n_a n_b -\rangle$. However, not all such transitions are allowed even when the resonance condition is satisfied. Owing to the parity symmetry of the two-boson-mode interaction Hamiltonian, only transitions to states with an even total number of excitations from the initial state $|00+\rangle$ are permitted. Consequently, the state $|00+\rangle$ can be resonantly coupled only to even-excitation states $|n_a n_b -\rangle$ satisfying $n_a+n_b\geq2$ and $n_a+n_b$ even. Representative examples of such multiquanta resonant transitions, namely $|00+\rangle\rightarrow|11-\rangle$ and $|00+\rangle\rightarrow|22-\rangle$, are shown in Figs.\,2(d) and 2(e) of the main text.

\section{Derivation of  effective multiquanta transition rates}
In the above section, we have derived the effective Hamiltonian
 \begin{align}\label{eqIII00}
H_{\rm eff}'&=\Delta_{a}a^{\dag}a+\omega_b b^\dag b+\Omega\tilde{\sigma}_z+\frac{1}{2}\lambda[(a^\dag b^\dag + a^\dag b + a b^\dag + ab)\tilde{\sigma}_z + (a^\dag b^\dag + a^\dag b - a b^\dag - ab)(\tilde{\sigma}^\dag - \tilde{\sigma})],
\end{align}
where the first three terms constitute the free Hamiltonian $H_0$, while the remaining terms form the interaction Hamiltonian 
 \begin{align}\label{eqIII00-1}
V=\lambda[(a^\dag b^\dag + a^\dag b + a b^\dag + ab)\tilde{\sigma}_z + (a^\dag b^\dag + a^\dag b - a b^\dag - ab)(\tilde{\sigma}^\dag - \tilde{\sigma})]/2,
\end{align}
which is responsible for inducing state transitions. To derive the multiquanta effective transition rates, we treat the interaction Hamiltonian $V$ as a  perturbation under the condition $\lambda\ll\Delta_a+\omega_b,2\Omega$.  According to the perturbation theory, the transition rate between an initial sate $|i\rangle$ and a finial state $|f\rangle$ can be given by 
\begin{align}\label{eqIII01}
&W_{|i\rangle\to|f\rangle}^{\rm eff} =2\pi |V^{\rm eff}|^2 \delta(E_{|f\rangle}-E_{|i\rangle}),
\end{align}
where $E_{|i\rangle}$ and $E_{|f\rangle}$ denote the energies of the initial state $|i\rangle$ and final state $|f\rangle$, respectively,  and $V^{\rm eff}$ is the effective coupling strength connecting the initial state $|i\rangle$ and final state $|f\rangle$. For resonant transition from $|i\rangle$ to $|f\rangle$, the transition rate $W_{|i\rangle\to|f\rangle}^{\rm eff}$ is determined predominantly by the effective coupling strength$V^{\rm eff}$:  a larger effective coupling leads to a higher transition rate. 

If the initial and final states are directly coupled through a first-order perturbation process, the  effective coupling strength is simply given by 
\begin{align}\label{eqIII01-1}
V^{\rm eff}=V_{fi}=\langle f | V | i \rangle.
\end{align}
If the coupling occurs only via second- or third-order perturbation processes, the corresponding effective coupling strengths are determined by high-order perturbation theory, involving one or more intermediate states. For the second- or third-order processes, the corresponding effective coupling strengths are given, respectively, by\,\cite{Ma2015Law,Garziano2016MS}
\begin{align}\label{eqIII02}
&V^{\rm eff}=\sum_{n} \frac{\langle f | V | n \rangle\langle n | V | i \rangle}{E_{|i\rangle}-E_{|n\rangle}}, 
\end{align}
and
\begin{align}\label{eqIII02-1}
V^{\rm eff}=\sum_{m,n} \frac{\langle f | V | n \rangle\langle n | V | m \rangle\langle m | V | i \rangle}{(E_{|i\rangle}-E_{|n\rangle}) (E_{|i\rangle}-E_{|m\rangle})},
\end{align}
where the states $|n\rangle$ and $|m\rangle$ are intermediate states involved in the transition pathways. More generally, if the initial and final states are connected only through higher-order perturbative processes, the effective coupling strength $V^{\rm eff}$ can be obtained in an analogous manner. The resulting effective interaction between the two resonant states can therefore be described by the reduced two-level Hamiltonian of the form
\begin{align}\label{eqIII03}
&H_{\rm eff}''=V^{\rm eff}|f\rangle \langle i | + {\rm h.c.}.
\end{align}

For the resonant transition from $|00+\rangle$ to $|11-\rangle$, the process can be induced directly by the tripartite interaction and does not involve any intermediate states, i.e., $a^\dag b^\dag \tilde{\sigma} |00+\rangle \to |11-\rangle$, then the resulting effective coupling strength can be obtained from first-order perturbation theory and is given by
\begin{align}\label{eqIII04}
V_{11}^{\rm eff}&=\langle 11- | V | 00+ \rangle\nonumber\\
& = \langle 11- | \frac{1}{2}\lambda[(a^\dag b^\dag + a^\dag b + a b^\dag + ab)\tilde{\sigma}_z + (a^\dag b^\dag + a^\dag b - a b^\dag - ab)(\tilde{\sigma}^\dag - \tilde{\sigma})] | 00+ \rangle\nonumber\\
&=\frac{1}{2}\lambda, 
\end{align}
where  $V_{11}^{\rm eff}\propto \lambda$, since the transition occurs directly without involving any intermediate states. Thus, the effective multiquanta  transition rate from  $|00+\rangle$ to $|11-\rangle$ is
\begin{align}\label{eqIII04-1}
W_{11}^{\rm eff} =2\pi |V_{11}^{\rm eff}|=\frac{1}{2}\pi \lambda^2.
\end{align}

In contrast,  the resonant transition from $|00+\rangle$ to $|22-\rangle$ occurs predominantly through two second-order transition pathways, i.e., $|00+\rangle\to|11+\rangle\to|22-\rangle$ and $|00+\rangle\to|11-\rangle\to|22-\rangle$, each involving an intermediate state. Consequently,  the effective coupling strength for this process needs to be evaluated using second-order perturbation theory. The resulting effective coupling is given by the sum of the contributions from both transition pathways, with 
\begin{align}\label{eqIII05}
V_{22}^{\rm eff}&=\sum_{n} \frac{\langle f | V | n \rangle\langle n | V | i \rangle}{E_{|i\rangle}-E_{|n\rangle}}\nonumber\\
&=\frac{\langle 22- | V | 11+ \rangle\langle 11+ | V | 00+ \rangle}{E_{|00+\rangle}-E_{|11+\rangle}}
+\frac{\langle 22- | V | 11- \rangle\langle 11- | V | 00+ \rangle}{E_{|00+\rangle}-E_{|11-\rangle}}\,\nonumber\\
&=\frac{\lambda^2}{2(\Delta_{a}+\omega_b)}+\frac{\lambda^2}{2[2\Omega-(\Delta_{a}+\omega_b)]}=\frac{\lambda^2}{2} \left[ \frac{1}{\Delta_{a}+\omega_b} +\frac{1}{2\Omega-(\Delta_{a}+\omega_b)} \right],
\end{align}
where  $V_{22}^{\rm eff}\propto \lambda^2$, since each dominant transition pathway involves one intermediate state. The effective multiquanta  transition rate from  $|00+\rangle$ to $|22-\rangle$ can thus be given by
\begin{align}\label{eqIII05-1}
W_{22}^{\rm eff} =2\pi |V_{22}^{\rm eff}|=\frac{1}{2}\pi\lambda^4\left(\frac{1}{2\Omega-\Delta_{ap}-\omega_b}+\frac{1}{\Delta_{ap}+\omega_b}\right)^2.
\end{align}
Figures 3(c) and 3(d) of the main text  compare the effective coupling strength $W_{11}^{\rm eff}$ (and $ W_{22}^{\rm eff}$) between the states $|00+\rangle$ and $|11-\rangle$ (and $|22-\rangle$) obtained from the analytical expression Eqs.\,(\ref{eqIII04-1}) [and Eqs.\,(\ref{eqIII05-1})] via the perturbation theory with results from the full-numerical simulations. The excellent agreement  between the two confirms the validity of our approximation.  

\end{document}